\documentclass[conference]{IEEEtran}

\usepackage{amsmath}   
\usepackage[dvips]{color}

\usepackage[pdftex]{graphicx}  
\usepackage{amssymb}
\usepackage{graphicx}
\usepackage{algorithm}
\usepackage{epstopdf}
\usepackage{color}
\usepackage{url}

\newtheorem{theorem}{Theorem}
\newtheorem{example}{Example}
\newtheorem{remark}{Remark}

\newtheorem{claim}{Claim}
\newtheorem{definition}{Definition}

\begin{document}

\title{Quorum Sensing for Regenerating Codes in Distributed Storage}


\author{\IEEEauthorblockN{Mit Sheth}
\IEEEauthorblockA{Sapient-Nitro\\Gurgaon, Delhi,\\ India}
\and
\IEEEauthorblockN{Krishna Gopal Benerjee}
\IEEEauthorblockA{Dhirubhai Ambani Institute of\\ Information and Communication Technology\\
Gandhinagar, Gujarat, 382007 India}
\and
\IEEEauthorblockN{Manish K. Gupta}
\IEEEauthorblockA{Dhirubhai Ambani Institute of \\Information and Communication Technology\\
Gandhinagar, Gujarat, 382007 India\\Email: m.k.gupta@ieee.org}
}

%

\maketitle

\begin{abstract}
Distributed storage systems with replication are well known for storing 
large amount of data. A large number of replication is done in order to provide
reliability. This makes the system expensive. Various methods have been proposed over 
time to reduce the degree of replication and yet provide same level of reliability. One recently suggested 
scheme is of Regenerating codes, where a file is divided in to parts which are then processed
by a coding mechanism and network coding to provide large number of
parts. These are stored at various nodes with more than one part at each
node. These codes can generate  whole file and can repair a failed
node by contacting some out of total existing nodes. This property
ensures reliability in case of node failure and uses
clever replication. This also optimizes bandwidth usage. 
In a practical scenario, the original file will be
read and updated many times. With every update, we will have to update
the data stored at many nodes. Handling multiple requests at the same
time will bring a lot of complexity. Reading and writing or multiple
writing on the same data at the same time should also be prevented. In
this paper, we propose an algorithm that manages and executes all the requests
from the users which reduces the update complexity. We also try to keep an adequate 
amount of availability at the same time. We use a voting based mechanism
and form read, write and repair quorums. We have also done probabilistic analysis of regenerating codes.
\end{abstract}


%
\IEEEpeerreviewmaketitle

\section{Introduction}
Cloud data storage is a challenging field posing many new problems  for researchers since new companies such as Facebook and Google  etc want to store several  terabytes of data of the 
users and still provide the smooth user experience of services. Many companies such as Amazon etc. are providing cloud data storage services and in doing so one has to solve multi-dimensional 
optimization problem. For such a cloud storage, in order to  make the data reliable one can use the classical  technique of replication, in which the data on the nodes is replicated many times. 
This technique of storing data is very easy but may consume a lot of space. It is used in popular distributed systems~\cite{10.1109/TPDS.2012.115},  
where data is divided into parts, and stored at various nodes. One advantage of distributed storage is reliability.
Even if one of the node gets corrupted or fails, then the other nodes still
remain alive. Most modern day distributed systems use replication \cite{10.1109/TPDS.2012.115,Huang:2012:ECW:2342821.2342823,Ghemawat03}. 
While using distributed systems, there is a high chance for the nodes to get damaged or to leave after some amount of time.
That new node will be directly copied from its replication as per the current storing scenario. The amount of space consumed in storing
the data in replicated manner is huge, for example, Gmail uses around $21$ times replication \cite{Ghemawat03}. 
One aim to use the already existing nodes to make the new node instead of the replicated nodes and thereby reduce the storage consumption.
In such systems  a file of size $M$ is divided into $n$ parts with each part of size $\frac{M}{n}$ bytes and stored 
at different nodes in such a way that any node can be reconstructed using a subset of total $n$ nodes.
These codes are called \textit{Erasure codes} \cite{4215814}. \textit{Maximum distance
separable(MDS)} erasure codes described in \cite{5709963}, \cite{4215814} are optimal erasure codes and can generate
a node using $k$ nodes where $k<n$. The main concern using MDS
codes is the bandwidth consumption as compared to replication. We need to download data from $k$ nodes in MDS 
codes instead of just one in replication.  In a seminal paper Dimakis et al. \cite{4215814} has shown that repair bandwidth can be reduced if we apply network coding~\cite{850663} on parts of the 
file and store $\alpha$ such linearly combined parts on every node.  In order to repair a failed node, we now
download data only from $d$ nodes instead of $k$, where $d<n$,
and from each node instead of downloading the whole data, we download
only $\beta$ packets out of $\alpha$, where $\beta<\alpha$.  These
codes are called \textit{Regenerating codes} \cite{4215814}. The repair bandwidth for the code is $d \beta$.
There is well known tradeoff between storage and repair bandwidth ~\cite{4215814}.
These codes can decrease the data that needs to
be downloaded from the nodes to repair or construct the new node. 
Each file is divided, encoded, and stored at various nodes. Considering
a practical scenario, there will be constant requests to read or write
the file. Some nodes might fail in between, so there will be
repair procedures going on. Any small change in the file will change
its associated parts at various nodes. There has to be a
proper method to do this, so that the multiple read's, repair's and
write's on the parts of the file are controlled in a proper way and
do not interfere with each other creating wrong results. Also this method
should not decrease the availability of the nodes and should
be able to perform many requests in parallel. This motivated us to
develop an algorithm by which the read, write and repair requests
are scheduled in a way such that we do not encounter any inconsistency
in the results and get the fastest results possible by processing
as many requests as possible in parallel. We use the basic concept
of simple voting mechanism with coding as in \cite{372774} where each node
is given one vote to describe its state. 

In this paper we give an algorithm based on quorum sensing
to carry out the operations of read, write and repair on the nodes
in a way such that the availability of the nodes also remain high
and that the operations are properly ordered. The algorithm is generic and can be applied to 
any variant of regenerating codes.
The rest of the paper is organized as follows. Background material  and origin of the problem is given in Section $2.$ 
Our proposed algorithm is described in Section $3$, Section $4$ discusses the Hadoop implementation and Section $5$ gives the probabilistic analysis of regenerating codes. 
Finally Section $6$ concludes the paper with some general remarks. 
\section{Preliminaries}
When we store data in a distributed system, we have to coordinate the access to the data on various nodes in such a way such that the operations are fast and their results are
free from error.  One has to optimize two main parameters viz. update complexity and repair bandwidth \cite{5709963}. 
Network coding \cite{850663} 
has shown a way to reduce the repair bandwidth and provide better reliability \cite{5709963}. In network coding, 
one performs mathematical operations on packets instead of just forwarding them. On the other hand 
quorum sensing has been used by researchers \cite{372774}  to give efficient algorithms that reduces 
update complexity in such a scenario.  One kind of voting mechanism, suggested by Gifford \cite{Gifford79weightedvoting}, assigns 
votes to the various  nodes and they communicate with each other using these votes. The read and write quorums are defined
to provide a read-write and a write-write exclusion. The read quorum
$r$ and write quorum $w$ are such that $r+w>N$ and $2*w>N$, where
$N$ is the total number of votes assigned to all nodes. An approach
called simple voting with coding (SVWC), as described in \cite{372774},
defines read and write quorums for a replication scheme using
coding to store files. The bounds on read and write quorum values
were studied in this paper. These bounds were calculated in a way
that you need at least votes equal to the lower bound to complete
the read/write quorum and if equal to upper bound will always complete
the read/write quorum. We use the basic concept from these algorithms
about how votes can be used in managing a big network of nodes and
communications with them.  Throughout the whole discussion we assume
that no votes are lost in the process and the server can fail only
if it is stopped voluntarily. We also assume that all the votes of one request
move at same speed in the network. 

\section{Proposed Algorithm}
In this section, we consider a code with MDS property and discuss the basic file operations in a practical scenario. 
We define various terms as we go on describing the algorithm. We take specific cases in the beginning and generalize them one after other. Each time we 
generalize by some amount, we derive new read, write and repair quorums. In the algorithm we design conditions that will prevent both read-write and write-write 
from executing at the same node at the same time.    
Consider a Maximum Distance Separable (MDS) code for a file $F$. The file is
initially divided into $k$ parts called as native chunks. Now, these
$k$ chunks are encoded by linear combination to form $n$ code chunks,
with $n > k$, and stored at $N$ nodes. Considering the MDS property,
any $k$ nodes can be contacted to reconstruct the whole file again.
The maximum number of failures which can be tolerated by the system
is $n-k$. Every node in the cloud stores $\alpha$ chunks making
$n\alpha$ chunks in total. We can contact any $d$ nodes and download
$\beta$ out of $\alpha$ packets from each node in order
to construct a new node which can replace the corrupted or failed
node. For any given regenerating code, let us say any change in file
$F$ updates $q$ chunks in total. There are three operations that
can be executed on the file. 
\begin{enumerate}
\item Read/download/reconstruct the whole file- There will be a lot of users
who would perform these requests. In this case, we will
have to generate the complete file, and to do that, we will have to
contact any $k$ nodes out of total $N$ nodes. 
\item Write/update- Whenever any file is changed, the corresponding data
in the nodes will also change. There will again be a lot of users
who would try to perform simultaneous write requests, which will update
the information stored in the nodes.
\item Node Repair- If a node gets corrupted or fails, we need to generate
a new similar node. For this, we need to connect to $d$ different
nodes and download $\beta$ packets from each of them. The node making
this request might be a specific node who is given the task of node
repair. 
\end{enumerate}
We  have divided the user requests into above three types. Everything
can happen in a smoothly if requests do not interfere with
each other. In a practical scenario, there will be constantly two or more requests which might be requested
at the same time. Such simultaneous requests might result in some
problem. There would be some inconsistency when we are reading from
a certain node and suddenly a write operation tries to write on it.
One way to operate and avoid inconsistency is to keep the incoming
operations waiting and that write operation will keep trying till
the read is completed. Following this approach, if another read
request comes after the write request, it will also be put in the
waiting queue behind write. There would be a lot of polling which
is a waste of resources because two reads can occur at a same time.
There will be many such requests trying to read or write a large number
of nodes. If these requests are not properly controlled and managed,
the whole network might get congested and may take too much time to
process user requests. As a solution to the given problem,
we propose a \textit{voting-locking} approach to manage such large
number of requests efficiently. We define quorum requirements to achieve this. 
\begin{definition}
The read, write and repair quorums are defined as the minimum number of votes required to
initiate the execution of that request. We denote the read, write
and repair quorum by $r$ , $w$ and $rep$. 
\end{definition}
Our main concern is to define a systematic approach which will try to reduce as many inconsistencies
as possible and give the result of the requests in least possible
time. Following are the six possible simultaneous request cases. Even if the simultaneous requests are more than two they will be made 
up of any such two pairs of requests only.
\begin{enumerate}
\item read-read 
\item read-write 
\item read-repair
\item write-write
\item write-repair
\item repair-repair 
\end{enumerate}

In order to describe the algorithm, let us first
take some assumptions and derive the read, write and repair quorums under these assumptions. We then gradually go on
generalizing after removing the assumptions one by one. 
\subsection{Single chunk node with simultaneous updates}
In this subsection we find out the quorum values for single chunk node with simultaneous updates possible. More precisely, 
we have the following:
\begin{enumerate}
\item Each node has only one code chunk stored. So number of code chunks
and number of cloud nodes are same i.e. $\alpha=1$, $n$$=$$N$
\item Any update in file $F$ changes every code chunk associated with it.
So as per assumption , data in every node changes with any change
in the file. 
\end{enumerate}
Under these two assumptions, it is easy to see that $N=n=q$, so we will
call one code chunk as one node for now. With these assumptions, let
us see which requests can execute simultaneously. 
\begin{itemize}
\item read-read - Yes. Two simultaneous reads will never be a problem. They
do not create any inconsistency.
\item read-write- No. As every write updates every node currently, there
will be nodes on which both operations would be executed simultaneously.
This will bring an inconsistency in read operation. 
\item read-repair- Yes, because repair is just another kind of a read request
contacting lesser nodes.
\item write-write- No. Simultaneous write's will definitely create inconsistency
in as both of them will try to update every node
at the same time giving incorrect results in the end.
\item write-repair- No. It is another kind of read-write repair. Even though
the number of nodes that we are contacting for repair operation is
less, the write operation is updating all the nodes and so there will
definitely be some common nodes and so the inconsistency.
\item repair-repair- Yes. It is an another kind of read-read case and so
it is possible. 
\end{itemize}
We define a unique \textit{vote-bit} and \textit{lock-bit} of one
bit on every chunk. When the \textit{vote bit} is 1, the node can
send votes and when it is 0, they cannot. \textit{Lock-bit} for a
node is a one bit value describing what lock is on it, read lock or
write lock. If the \textit{lock bit} is 0 then there is no lock on
it. We call the server a \textit{super-user} from which the read,
write and repair requests will be executed and end nodes as \textit{end
users} from where the requests are coming. This is shown in Figure
$1$. 
\begin{figure}
\includegraphics[scale=0.6]{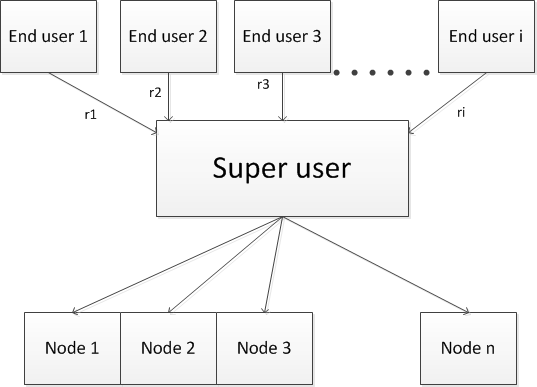}
\textit{\caption{This depicts a server with end users and nodes. The end users are those who send requests of type read, write, repair to the
super user. The nodes are the storage blocks where the data of file
is stored. After receiving various kind of requests from the end users,
the super user tries to communicate with the nodes to find out the
nodes available for performing requests on them.}
}
\end{figure}
Whenever the \textit{super user} gets any kind of request, it will
check if the nodes are free to execute that request on them. We will
discuss later the pattern in which the \textit{super-user} will send
requests. We now define two kinds of locks.
\begin{enumerate}
\item Read lock- Whenever any node is free to be read, we apply a
\textit{read lock} on it keeping its \textit{vote bit} unchanged i.e.
equal to 1 and \textit{lock bit} to 1. When the \textit{read
lock} is set, it cannot surpass write requests but can allow multiple
read and repair requests. Once \textit{read lock} is set, the \textit{super
user} starts reading from that node and when the read operation is
done, the \textit{super user} node frees it from the \textit{read
lock} and the \textit{lock bit} is changed to 0.
\item Write lock- Whenever any node is free to get updated, we try to form
a \textit{write lock} on it after changing its \textit{vote bit} to
0 and \textit{lock bit} to 2. When the \textit{write lock} is set,
it cannot allow any kind of further requests and the \textit{super user} starts updating that node. When
the write operation is done, the super user node frees it from the
\textit{write lock} and changes its \textit{vote bit} to 1 again and
\textit{lock bit} to 0.
\end{enumerate}
We define a \textit{lock table} at the \textit{super user},
which enlists if there is any lock at a node, and if there
is one, which lock is it. When the \textit{super user} gets a request
from \textit{end users}, it tries to know the availability of nodes
for that request by sending the request to the nodes. Every request
from the super user to the nodes has a unique \textit{request id}.
Nodes will be available if there is no lock on them, and if there
is a lock, then the availability will depend on the type of
lock on them. If they are available for a certain request, they
will send a vote to the \textit{super user} in form of a two bit \textit{(node
number,request id)} number only if their \textit{vote bit} is 1. When this
vote reaches \textit{super user}, it checks in the lock table if that
node can be granted that lock. If it can be granted, then it updates
its \textit{lock table} with the corresponding lock on the request
id, and then updates the \textit{vote bit} and \textit{lock bit} on
the node, if required. It would not be necessary to update the \textit{vote
bit} and \textit{lock bit} if the node on which the request was granted
was already locked, because these requests will be read-read,
read-repair or repair-repair and all these have same \textit{vote
bit} and \textit{lock bit}. The node structure can be seen in Figure
2 and the \textit{lock table} can be seen in Figure $3.$ 
\begin{remark}
We use a read lock for both, the read and repair request and only
a read lock can be granted in the presence of another read lock but
no other lock can be granted in the presence of a write lock. A read operation can be done from any set 
of nodes, so we are not restricted to read from specific nodes after putting a \textit{read
lock} on them. However, write's for a request are specific. They need
to be done on specific set of nodes. So, if a write lock is not obtained
on specific chunks, then it waits for the current lock on it to release.
If one more write request arrives for the same chunk, then it will
also go in the waiting list.
\end{remark}
\begin{figure}
\[
\begin{array}{c}
\includegraphics[scale=0.7]{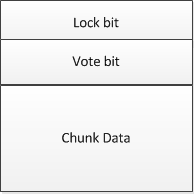}
\end{array}
\]
\caption{This diagram represents the node structure of every node and lock
table which is stored at the super user. The structure of the every
node consist of lock bit, vote bit, and the information to be stored
in it. Lock bit gives the information about the lock on the node and
vote bit and its tells if that node has the ability to send votes
or not.}
\end{figure}
\begin{claim}
Quorum values for single chunk node with simultaneous updates are given below.
\begin{enumerate}
\item $r=k$, because we need at least $k$ nodes from out of $N$ to read
to download/read the file.
\item $w=n=N=q$, because every write is updating every node, so it requires
all nodes to be available for writing at a time. 
\item $rep=d$, because it requires any $d$ nodes to read from to reconstruct
the failed node. 
\end{enumerate}
\end{claim}
We take the NCCloud system \cite{nccloud012} to explain this scenario. If we
look at the encoding mechanism of NCCloud from $k$ native chunks to
$n$ code chunks, we will see that each code chunk is dependent on
every native chunk. Thus, even a small change in the file will change
all the $n$ chunks. Now, if the $8$ code chunks of NCCloud were stored
on $8$ nodes instead of $4,$ then this would be a perfect example as per
our assumptions. So, assuming $N=8,$ and each chunk stored on individual
node, we can say that for this system, the read and write quorums
will be $r=2,$ and $w=8.$
\begin{example}
Consider a different scenario with $N$ =10, $r$=6 and $w$=6. What
if 5 out of 10 nodes allow read operation on them and other 5 allows
write operation on them. Neither of the quorums will get satisfied
and both of them will wait for one extra vote which they will never
get in this condition. This is a \textit{deadlock situation}. In order
to solve this we define two terms, \textit{timeout time} $t_{0}$
and a \textit{request queue}. A request queue is a queue which distributes
the incoming requests in to slots, where all the requests of one slot
are handled together and every request has its own priority. The priority
for every request {[}$read{}_{0}$, $read{}_{1}$,$write{}_{0}$,
$write{}_{1}$…$rep{}_{0}$, $rep{}_{1}$…{]} is zero initially. The
requests with higher priority than others in the same slot are handled
first. The division of the slots can be done on the basis of time
or number of requests i.e. one slot can be of all the requests that
arrived in time $t$ or one slot can be a slot till $r{}_{s}$ requests
arrive at \textit{super user}. Here $r{}_{s}$can be a read or write
or repair request. The \textit{request queue} is handled at the \textit{super-user}
level and once a request comes in the running slot, it does not wait
for the slot to fill, it just starts asking for votes. We define
\textit{request-vote ratio} for a request as the ratio of the votes
obtained to the quorum votes needed by that request. If the quorum
requirements in the running slot are not completed for any of the
requests for a continuous time of $t{}_{0}$, then we assume it as
a \textit{dead lock situation} and pick one-fourth of the requests
with minimum \textit{request-vote ratio}, decrease their priority
and then release their acquired locks on nodes. If the situation remains
same even after an another interval of $t_{0}$, this is repeated
again and then the priority of the requests whose priority were reduced
earlier will be further reduced. We now further limit our assumptions
as below.
\end{example}
\subsection{\noindent \textit{Single chunk node with relative updates} }
We now remove the \textit{assumption (2)} that we took in \textit{Section A}
and say that any update in file $F$ changes $q$ code chunks out
of $n$. We are now left with following assumption.
\begin{enumerate}
\item Each node has only one code chunk stored. So number of code chunks
and number of cloud nodes are same i.e. $\alpha=1$, $n$$=$$N$
\end{enumerate}
Let us now see which of the consecutive requests can hold. 
\begin{itemize}
\item read-read - Yes, because two simultaneous reads are never a problem.
\item read-write- May be. Here every write updates $q$ node(chunks), hence
there might be a possibility when there will be no common nodes among
both operations.
\item read-repair- Yes, because repair is just another kind of a read request
contacting lesser nodes.
\item write-write- May be. Simultaneous write will definitely create inconsistency
in write operations but if both the write's are writing on completely
different set of $q$ nodes, then this would be possible.
\item write-repair- Yes, if the intersection set of nodes for both requests
is zero.
\item repair-repair- Yes, this is also another kind of read-read and so
it is possible. 
\end{itemize}
In this case,  quorums are
\begin{claim}
Quorum values for single chunk node with relative updates are given as:
\begin{enumerate}
\item $r=k$, because we need at least k nodes to read from to download/read
the file. 
\item $w=q$, because every write is updating q nodes at a time 
\item $rep=d$, because it requires any $d$ nodes to read from to reconstruct
the failed node. 
\end{enumerate}
\end{claim}
\subsection{Multiple chunk node with relative updates}
Let us now remove the assumption in \textit{section B} also and
say that each node has $\alpha$ code chunks stored in it. Every write
updates $q$ out of $N\alpha$ code chunks. We cannot take votes from
individual chunks, we have to take them from nodes only. We give $\alpha$
votes to each node instead of one. Every vote is distinct, even on
the same node. Node $i$ will have votes from $\alpha{}_{i1},\alpha{}_{i2}....\alpha_{i\alpha}$
for its $\alpha$ number of packets, where 1$\leq$$i$$\leq N$.
We will now treat one node and one chunk differently. This can be
seen in Figure $3.$ The conditions of the consecutive requests will be same as described
under \textit{Section B}.
\begin{figure}
\[
\begin{array}{c}
\includegraphics[scale=0.6]{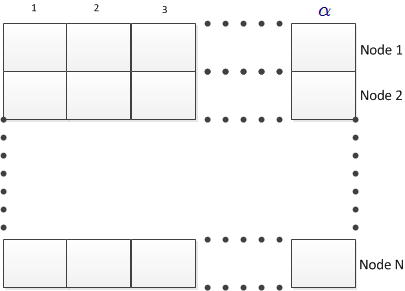} \\ 
\includegraphics[scale=0.6]{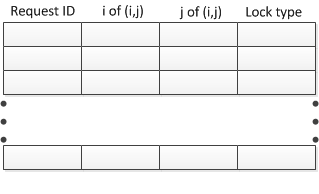}
\end{array}
\]
\caption{This shows how chunks are distributed in nodes and the way in which
each chunk can be uniquely identified. Lock table consists of four
coloums where request id describes the unique request id of every
request. i and j describe the node number and lock type describes
the lock.}
\end{figure}
Due to generalization, we do a little modification in the mechanism
by which the availability of nodes was decided. Whenever the \textit{super
user} asks the nodes for acceptance of request, we check \textit{vote
bit} and \textit{lock bit} of every packet of every node. If the chunk
is able to allow that request, it sends a 3 bit vote $[i,j,requestid]$
where $i$=node number and $j$= chunk number of that node and requested
is the unique id of the request that came to that chunk. Quorum values
in this case are given below.
\begin{claim}
Quorum values for multiple chunk node with relative updates are given as:
\begin{enumerate}
\item $r=\sum\alpha_{ij}$, where $i=$any $k$ values from $1$ to $N$
and $j=\alpha$.
\item $w=q$, because every write is updating q chunks at a time 
\item $rep=\sum\alpha_{ij}$, where $i$$=$ any $d$ values from $1$ to
$N$ and $j=$any $\beta$ values from $1$ to $\alpha$.
\end{enumerate}
\end{claim}
and $\sum\alpha_{ij}$indicates the total number of votes, with $\alpha_{ij}$as
one vote.
\begin{example}
Consider a case when we try to perform a read request and a write
request and both have two chunks common in their operations as explained
below. If $r=4$, $w=4$ and we have 6 chunks. Suppose the vote for
both read and write lock goes to the \textit{super user} but \textit{read
lock} is formed first and then the \textit{write lock}. So currently
we have \textit{read locks} on 4 chunks, write lock on 2 chunks and
a waiting \textit{write lock} on 2 chunks that are being read currently.
Meanwhile, suppose the other 2 chunks get updated and lock is removed
from them. Now, suppose a new read request is made and it tries to
read from the two recently updated chunks and the two on which the
\textit{write locks} are waiting. Now, this is not supposed to happen,
we are trying to read from the chunks that got updated and those who
are waiting to be. This might result into an incorrect read, so we
add one more condition. We are not allowed to perform any new request
$r{}_{i}$on a chunk with a waiting write/read/repair lock of request
$r{}_{k}$on it, when operations of request $r{}_{i}$ are supposed
to involve both, the chunks updated by and the chunks with waiting
locks from the same request $r{}_{k}$.
\end{example}
\begin{figure}
\includegraphics[scale=0.7]{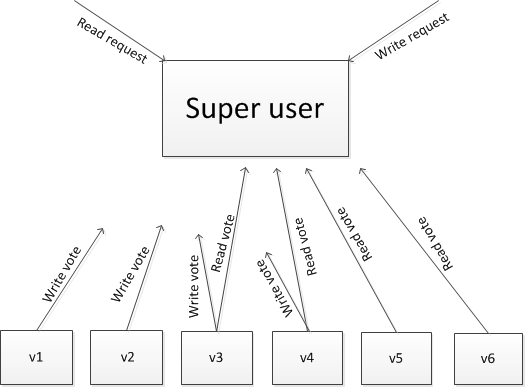}
\caption{This situation depicts the case of Example $3$ where the super user has
got two request read and write. We have six nodes $v_1, v_2, v_3, v_4, v_5, v_6.$ Four nodes sends the approval for
read request by sending the read vote to super user. Before the lock
table gets updated and lock bit and vote bit of these four nodes update
two of that four nodes come across write request and sends approval votes
for write request also. Another two different nodes, $v_1$ and $v_2$ also
sends vote for write request as shown. However as the read votes are
received first, a read lock is formed first on $v_3$ and $v_4$. }
\end{figure}
So, as per the Figure $4,$ read lock will be enabled on nodes $v_4, v_5$ and
$v_6,$ write lock on nodes $v_1$ and $v_2$ and a waiting write lock on node $v_3$.
If the operation of write on $v_1$ and $v_2$ is completed then they would
be freed from existing write locks. Suppose a new read request comes
with the current read request still going on. This new request will
be able to form a lock on $v_1$ and $v_2$ easily. As per our current
algorithm, $v_3$ and $v_4$ should also be allowed to be read under the
new read request, but if we do so, the constructed file might be incorrect
as the MDS property might not hold for nodes with only half of an
operation done. Thus, we restrict ourselves from choosing one or more
from the both the sets $\{v_1, v_2\},\{v_3, v_4\}.$ We can
still select one or more of $\{v_1, v_2\}$ with some other nodes
or one or more of $\{v_3, v_4\}$ with other nodes. This will prevent
from any possible inconsistency in the network.
Let us discuss the way in which the \textit{super user} should send
the request for votes to the nodes. There should be some proper way
for the \textit{super-user} to carry out the process efficiently.
Multicasting is one way but it will increase the overhead by a large
amount. We try form some groups which can reduce the
overhead. After we form groups, instead of multicasting, write requests are sent group wise and read requests are sent depending upon the existing write requests to create least possible interference. One such approach is as follows. We have $k$ native chunks
and any change in the file will be replicated in any of these $k$
native chunks. We can look at encoding mechanism of constructing $n$ coded chunks from $k$ native chunks 
and can exactly say which out of the $n$ coded chunks will change for a 
corresponding change in some native chunk. We can look further and know which $q$ chunks out of $\alpha N$ will change for a corresponding change in native chunk. There are $k$ native
chunks, so there will be $k$ groups overall. We make a
group of all the chunks which are changing due to change in some native
chunk. There will be $k$ groups formed each containing $q$ chunks.
We have a total of $N$ each with a capacity of $\alpha$ chunks. We
try to store the first group of $q$ chunks in first node. There wont
be any problem if $q<$$\alpha$ but if $q>\alpha$, then we store
the remaining $\alpha-q$ chunks on second node. We now
search other groups and find which group has highest number of chunks that are
common with the $\alpha-q$ chunks. We store all the chunks of that group other than the chunks that were common with $\alpha-q$ chunks on second node. We keep on making groups in this way till
every chunk gets placed somewhere. If a read request comes first, it is sent to random k nodes which are adjacent to one another. When a write request
comes, the super user first checks which group will be affected most
by that change and sends request to that group. We know the chunks that needs to be updated and the group associated. If that group corresponds
to nodes $N{}_{i}, N{}_{i+1}$, etc., the write request is sent those corresponding chunks on the corresponding nodes and the next incoming read request
 is sent to the $k$ nodes which has higher \textit{write distance}. 
\begin{definition}
The \textit{distance} between two nodes is equal to the number of hops from node to node in between those two nodes. 
\end{definition}
\begin{definition}
The  \textit{write distance} for a node is defined as the sum of the \textit{distance} between the that node and the nodes on which the write operation is being performed at the present time. 
\end{definition}
\begin{remark}
This \textit{write distance} for all the nodes should be kept regularly updated in a table called \textit{write distance table} and stored at \textit{super-user}. Whenever a read request comes, we pick out the $k$ nodes with highest \textit{write distance} and send read requests to them. This helps in keeping read and write operation as far as possible.
\end{remark}
\begin{example}
Consider a scenario with $N=5$ nodes with write operation being performed
on any number of chunks at nodes $v_1$ and $v_3$. The current
\textit{write distance} for nodes $v{}_{1}$,$v_{2}$,$v_{3}$,$v_{4}$
and $v_{5}$ will be $2, 2, 2, 4$ and $6$ respectively. In a case if we
have to pick any one node for the read request, we will pick the node
$v_{5}$ first and if we have to pick one more node then we will pick
node $v{}_{4}$ after $v_{5}$. 
\end{example}
\begin{algorithm}
\caption{Read Algorithm for any request $r_i$}
\begin{enumerate}
\item Divide in groups $G_1, G_2, \ldots G_k$ such that $|G_i|=N_i, 1 \leq i \leq k$
\item Send vote requests to all the chunks in the nodes $N_1, N_2, \ldots N_k$, 
with maximum write distance
\item Define \\ $V=\{\;\mbox{set of nodes whose all chunks have sent ready votes}\;\}$
\item if $|V|=k$, Form a read lock on these nodes and start read operation 
\item Else,  Form a read lock on the nodes in $V$
\item Do
\item Contact $k-n(V)$  out of $N-n(V)$ with highest write distances. 
\item Add those nodes to $V$ whose every chunk has sent read vote and form a read lock on 
these nodes. 
\item WHILE $(n(V)!=k)$ 
\item Contact these k nodes present in V to perform read operation. 
\end{enumerate}
\label {alg3}
\end{algorithm}
\begin{algorithm}
\caption{Write Algorithmfor any request $w_i$ }
\begin{enumerate}
\item Find the chunks which will be changed for the write request $w_i.$ 
\item Find the nodes on which these chunks are stored in groups. 
\item Send vote requests to all the chunks in the corresponding nodes. 
\item If any other locks are already present on the nodes, then wait for those locks to release. 
\end{enumerate}
\label {alg3}
\end{algorithm}
\section{Implementation in Hadoop}
We have discussed the algorithm. Let us try to apply this to current
storage systems like Hadoop. The \textit{Hadoop Data File Structure(HDFS)}
is comprised of interconnected clusters of data in it. Every cluster
has one single node called \textit{Name Node} and many number of servers.
Each server has \textit{Data Node} with it which contains the data.
Name Node is the one who handles all the read and write requests from
the HDFS clients. Data Nodes perform action on data as per the instructions
given by the Name Node. Data Nodes continuously loop and ask the name
node for instructions. Following changes in the Hadoop system are
proposed to make it more efficient.
\begin{enumerate}
\item The Name node of the each cluster should be modified to act like the
super user node described in our system. The data on the data nodes
will be stored in form of regenerating codes instead of normal distributed
storage. The benefits of storing as regenerating codes over normal
distributed storage have already been described. Once this is established
the Name node will work on its own and the data nodes wont have to
continuously ask the name node for instructions. This will reduce
the communication overhead in an interaction between name node and
data node. Millions of such interactions take place in small interval
of time. Due to this, even a very small improvement in the overhead
will lead to a countable decrease in the overall overhead.
\item Files can be stored on the clusters as per our group forming mechanism
with group $1$ parts stored on data node $1$ and so on. This will further
reduce the communication overhead.
\item By default the replication factor in Hadoop system is $3.$ However,
the replication factor can be increased by companies depending on
the data they want to store. We can make the system work with 2 times
replication using our algorithm with amost equal or more reliability
than with replication factor as 3. For this, we store the \textit{n}
parts of the file on all the clusters except the one where the file
is stored. We divide the $n$ parts into various clusters in the way
that every cluster has almost same and minimum number of parts. Our
system will be able to tolerate many failures, cluster failure or
data failure.
\end{enumerate}
\begin{example}
If the total number of clusters are $15$ and the vale of $n$ is $40,$ we
have to divide these $40$ parts in $14$ clusters. For distributing, we
put the first $14$ parts on $14$ clusters, then another $14$ parts on $14$
clusters making each cluster store $2$ parts. We are still left with
$12$ parts. We divide this $12$ parts on any $12$ clusters out of $14$ making
$12$ clusters to store $3$ parts each and $2$ clusters to store $2$ parts
each.
\end{example}  
\section{Probabilistic Analysis of Regenerating Codes}
In this section, we look at the second aspect of the paper. We analyze the incoming request, completing  request and calculate the probabilities of downloading a file and repairing a node in a regenerating code.
\subsection{Probability Calculation}
Let the probability to read and write a node be $p_r$ and $p_w$ respectively then it is clear that 
\begin{equation*}
\begin{split}
&P\left(\mbox{download a file} \right) = \left\{ \,
\begin{IEEEeqnarraybox}[][c]{l?s}
\IEEEstrut
0 & if $i<k;$ \\
\sum_{i=k}^N{N \choose i}{i \choose k} p_r^i\left(1-p_r\right)^{N-i} & if $i\geq k.$
\IEEEstrut
\end{IEEEeqnarraybox} 
\right. \\
&P\left(\mbox{repair a node} \right) = \left\{ \, 
\begin{IEEEeqnarraybox}[][c]{l?s}
\IEEEstrut
0 & if $i<d;$ \\
\sum_{i=d}^{N-1} {N \choose i}{i \choose d} p_r^i\left(1-p_r\right)^{N-i} & if $i\geq d.$
\IEEEstrut
\end{IEEEeqnarraybox}
\right.
\end{split}
\end{equation*}
\subsection{Analysis of Incoming Request}
For regenerating codes we can make the following assumptions.
\begin{enumerate}
	\item There are $n$ requests of some operations (either read or write) in the system associated with a chunk $(i,j)$ at time $t$ and the probability of exactly one arrival request is given by $\lambda \Delta t + O\left(\Delta t\right)$ during a small interval $\Delta t$ where $\lambda$ is arrival rate of request independent of $t$.
	\item $\Delta t$ is so small that the probability of arrival of more then one request is almost zero.
	\item The incoming  requests are independent of each other.
\end{enumerate}
Under these assumptions using the Queueing theory \cite{saaty1961elements}, we get
\begin{theorem}
If the arrival of requests are completely random, then the the number of arrival requests follows Poisson distribution in a fix time interval. 
\label {Queueing theory}
\end{theorem}
Because of read as well as write requests  satisfy the above three assumptions so we can say that the read requests and the write requests follows the Poisson distribution separately with request arrival rate $\lambda_r$ and $\lambda_w$ respectively.  Hence the probability of $m$ read requests associated with a chunk $(i,j)$ at time $t$ is given by  
\begin{equation*}
P_m\left(t \right) = \frac{\left(\lambda_r t\right)^me^{-\lambda_r t}}{m!}.
\end{equation*}
Similarly the probability of $l$  write requests associated with a chunk $(i,j)$ at time $t$ is given by 
\begin{equation*}
P_l\left(t \right) = \frac{\left(\lambda_w t\right)^le^{-\lambda_w t}}{l!}.
\end{equation*}
Now on a chunk $(i,j)$ writing and reading is not possible at the same time so we can say that  after finishing writing request the chunk resets its time clock to $t=0$ for rest of read requests.
Hence let $P_{0,m}\left(t\right)$ be the probability to perform $m$ read operations  at time $t$ and $P_{l,0}\left(T\right)$ be the probability to perform $l$ write operations at time $T$ then 
\begin{equation*}
P_{0,m}\left(t \right) = \frac{\left(\lambda_r t\right)^me^{-\lambda_r t}}{m!} \mbox{ and }P_{l,0}\left(T \right) = \frac{\left(\lambda_w T\right)^le^{-\lambda_w T}}{l!}.
\end{equation*}
If $P_{l,m}\left(T,t\right)$ is the probability to perform $m$ read operations at time $t$ after performing $l$ write operations at time $T$.  Then 
\begin{equation}
\begin{split}
P_{l,m}\left(T,t\right)= P_{l,0}\left(T\right)P_{0,m}\left(t\right) \\
 = \frac{\left(\lambda_w T\right)^le^{-\lambda_w T}}{l!}\frac{\left(\lambda_r t\right)^me^{-\lambda_r t}}{m!}.
\label{equ.1}
\end{split}
\end{equation}
\begin{remark}
Note that
\begin{equation*}
\begin{split}
&\sum_{l=0}^\infty \sum_{m=0}^\infty \frac{\left(\lambda_w T\right)^le^{-\lambda_w T}}{l!}.\frac{\left(\lambda_r t\right)^me^{-\lambda_r t}}{m!} = 1 \mbox{ and} \\
&\;\mbox{expectation of}\;P_{l,m}\left(T,t\right)\;\mbox{is}\; E\left(l,m\right)  = \left(\lambda_wT\right)\left(\lambda_rt\right).
\end{split}
\end{equation*}
Hence expected request for read and write for a chunk in $T+t$ time is $\left(\lambda_w\lambda_rTt\right)$.
\end{remark}
\subsection{Analysis of Completing Request}
\begin{figure}
\centering
\includegraphics[scale=0.50]{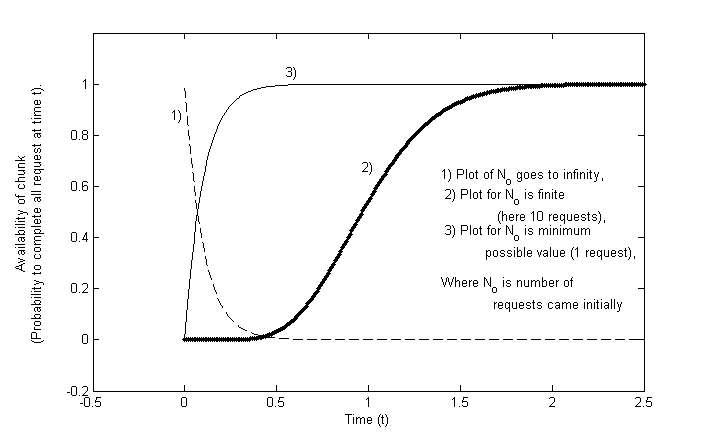}
\caption{Probability  $P_0(t)$  of completing read requests  with respect to time with $\mu_r$ = $10$ requests / unit time.}
\label{FlowGraph}
\end{figure}

For analyzing the availability of a chuck we need to analyze the requests (read or write) that has been completed.
Assume that  $N_o^{(r)}\left(N_o^{(w)}\right)$ requests are associated with the chunk $(i,j)$ for read (write) operations at time $t$ = 0 and no incoming requests are allowed from time $t$ = 0 to $t = t\left\{T\right\}$. 
Let the rate to complete the request be $\mu_r\left(\mu_w\right)$ per unit time. We  make the following assumptions 
\begin{enumerate}
	\item Probability of completing one read (write) request is $\mu_r\Delta t\left(\mu_w\Delta T\right)$.
	\item Probability of completing more then one request is zero.
	\item The completing  requests are independent of each other.
\end{enumerate}
Under the assumptions, by Queueing theory \cite{saaty1961elements}, the probability to have $m(l)$ read(write) request at time $t(T)$ i.e, $P_m\left(t\right)\left\{P_l\left(T\right)\right\}$ is given by
\begin{equation}
\begin{split}
&P_m\left(t\right)= \left\{ \,
\begin{IEEEeqnarraybox}[][c]{l?s}
\IEEEstrut
\frac{\left(\mu_r t\right)^{N_o^{(r)}-m}e^{-\mu_r t}}{\left(N_o^{(r)}-m\right)!} & if $m=1,2,...,N_o^{(r)};$ \\
1-\sum_{n=1}^{N_o^{(r)}}\frac{\left(\mu_r t\right)^{N_o^{(r)}-n}e^{-\mu_r t}}{\left(N_o^{(r)}-n\right)!} & if $m = 0.$
\IEEEstrut
\end{IEEEeqnarraybox}
\right. \\
& and \\
& P_l\left(	T\right)= \left\{ \,
\begin{IEEEeqnarraybox}[][c]{l?s}
\IEEEstrut
\frac{\left(\mu_w T\right)^{N_o^{(w)}-l}e^{-\mu_w T}}{\left(N_o^{(w)}-l\right)!} & if $l=1,2,...,N_o^{(w)};$ \\
1-\sum_{n=1}^{N_o^{(w)}}\frac{\left(\mu_w T\right)^{N_o^{(w)}-n}e^{-\mu_w T}}{\left(N_o^{(w)}-n\right)!} & if $l = 0.$
\IEEEstrut
\end{IEEEeqnarraybox}
\right.
\end{split}
\label{equ.2}
\end{equation}
Note that
\[
	P_0(t)  = 1-\sum_{n=1}^{N_o}\frac{\left(\mu_r t\right)^{N_o}e^{-\mu_r t}}{N_o!}.\;\mbox{Thus}\; \lim_{N_o\rightarrow \infty} P_0(t)  = e^{-\mu t}.
\]
Figure \ref{FlowGraph} plots the probability $P_0(t)$ of completing requests with respect to time $\mu_r$ = $10$ requests / unit time. Similar analysis can be done for write requests. 
\subsection{Analysis of Incoming and Completing Requests}
Let $\mu$ and $\lambda$ be the rates of completing and receiving requests in the system associated with the chunk. Then by queueing theory \cite{saaty1961elements} the probability of $n$ requests (read or write) in system is given by
\begin{equation}
P_n\left(t\right)= 
\begin{IEEEeqnarraybox}[][c]{l?s}
\IEEEstrut
\left(\frac{\lambda}{\mu}\right)^n\left(1 - \frac{\lambda}{\mu}\right), & $\left(\frac{\lambda}{\mu} < 1, n \geq 0\right)$. 
\IEEEstrut
\end{IEEEeqnarraybox}
\label{equ.4}
\end{equation}
\section{Conclusion}
We proposed an algorithm that can be used to manage the read, write and repair requests on data which is stored using regenerating codes. The algorithm proposed a solution to manage a large number of requests in a way such that they can be executed in least possible time. The algorithm also proposed a way to prevent inconsistencies that can happen due to read and write performed on the same node at the same time. It would be interesting in future to extend this algorithm further and solve other problems listed below. 
\begin{itemize}
\item Some addition to the algorithm can be done to manage the system in
case of loss of votes in the path and the case of sudden node failure
while performing operations on it. 
\item The algorithm can be extended to a large number of servers involved
in the network where single file will be connected to huge number
of servers.
\item Some other group formation technique can be formed which can decrease
the communication overhead in the process.
\item Cache technique can be applied to further decrease the communication
overhead.
\item The algorithm can be further extended and applied on \textit{Fractional Repetition (FR)} codes.
\end{itemize}
To analyze the requests associated with regenerating code we have used queueing theory.





%
%
\bibliographystyle{IEEEtranS}
\bibliography{biblo}

\end{document}